\documentclass[reprint,amsmath,amssymb,aps,prl]{revtex4-1}
\usepackage[dvips]{graphicx}
\usepackage{dcolumn}
\usepackage{bm}
\usepackage{float}
\usepackage{hyperref}
\usepackage{color}
\usepackage{cancel}
\usepackage{bmpsize}
\usepackage{amsmath}
\usepackage{amssymb}
\usepackage{natbib}

\newcommand{\bra}[1]{\left\langle #1\right|}
\newcommand{\ket}[1]{\left| #1\right\rangle}

\newcommand{\ketbra}[2]{\left| #1\right\rangle\!\left\langle#2\right|}

\newcommand{\be}[0]{\begin{equation}}
\newcommand{\ee}[0]{\end{equation}}

\newcommand{\tr}[0]{{\rm Tr}}

\newcommand{\lra}\simeq
\newcommand{\eeqref}[1]{Eq.~(\ref{#1})}
\newcommand{\braketop}[3]{\left\langle
#1\vphantom{#2#3}\right|\left.#2\vphantom{#1#3}\right|\left.#3\vphantom{#1#2}\right\rangle}

\definecolor{linkcolor}{rgb}{0.9,0,0}
\definecolor{citecolor}{rgb}{0,0.6,0}
\definecolor{urlcolor}{rgb}{0,0,1}
\hypersetup{
	colorlinks, linkcolor={linkcolor},
	citecolor={citecolor}, urlcolor={urlcolor}
}

\begin{document}

\graphicspath{{../}}
\title{Entanglement and teleportation\\ between polarization and wave-like encodings of an optical qubit}

\author{ Demid V. Sychev$^{1,4}$, Alexander E. Ulanov$^{1,2}$, Anastasia A. Pushkina$^{1,5}$, Egor Tiunov$^{1,2}$, Valery Novikov$^{1,2}$, and A. I. Lvovsky$^{1,3,5}$}

\affiliation{$^1$Russian Quantum Center, 100 Novaya St., Skolkovo, Moscow 143025, Russia}
\affiliation{$^2$Moscow Institute of Physics and Technology, 141700 Dolgoprudny, Russia}
\affiliation{$^3$P. N. Lebedev Physics Institute, Leninskiy prospect 53, Moscow 119991, Russia}
\affiliation{$^4$Moscow State Pedagogical University, Department of Theoretical Physics, M. Pirogovskaya Street 29, Moscow 119991, Russia}
\affiliation{$^5$Institute for Quantum Science and Technology, University of Calgary, Calgary AB T2N 1N4, Canada}

%Include the email address of the corresponding author here.
\email{LVOV@ucalgary.ca}
\date{\today}

\begin{abstract}
Light is an irreplaceable means of communication among various quantum information processing and storage devices. Due to their different physical nature, some of these devices couple more strongly to discrete, and some to continuous degrees of freedom of a quantum optical wave. It is therefore desirable to develop  a technological capability to interconvert quantum information encoded in these degrees of freedom. Here we generate and characterize an entangled state between a dual-rail (polarization-encoded) single-photon qubit and a qubit encoded as a superposition of opposite-amplitude coherent states. We furthermore demonstrate the application of this state as a resource for the interfacing of quantum information between these encodings. In particular, we show teleportation of a polarization qubit onto a freely propagating continuous-variable qubit. %Because these are the primary qubit encodings used for quantum optical information processing, the capabilities developed here open new horizons in this field, particularly in quantum communications.
\end{abstract}

\maketitle

\vspace{10 mm}

\textbf{Introduction.} 
Different physical systems with the potential for quantum processing and storage can be roughly classified into two categories. Some systems, such as single atoms, quantum dots, superconducting circuits or color centers, have non-equidistant energy level structures, from which one can select a pair of levels that can serve as a qubit. For other systems, e.g., atomic ensembles, optical or microwave cavities, and optomechanical membranes, the energy level structure is inherently equidistant, and therefore analogous to that of the harmonic oscillator. In these systems, it may be more beneficial to encode quantum information in continuous degrees of freedom, such as the position and momentum.

Because different quantum systems are more suitable for performing different tasks, a technology for coherent and loss-free exchange of quantum information among them is essential for efficient integrated quantum information processing \cite{Kur15}. A natural mediator for such exchange is the electromagnetic field, which is the only quantum system capable of carrying quantum information over significant distances. Fortunately, this field is capable of coupling efficiently to both qubit-like and harmonic-oscillator-like systems through its own discrete- \cite{Kok} and continuous-variable (CV) \cite{Braunstein} degrees of freedom.

The most common discrete-variable (DV) approach to encoding quantum information in an optical wave is the dual-rail qubit: a single photon occupying one of two orthogonal modes corresponds to logical $0$ or $1$. These two modes can correspond, for example, to the horizontal $\ket H$ and vertical $\ket V$ polarizations. In the CV domain, a qubit can be encoded as a superposition of coherent fields of opposite phases, $\ket{\gamma}$ and $\ket{-\gamma}$, with the amplitude $\gamma$ being high enough to ensure sufficient orthogonality of these states \cite{Jeong, Ralph2003}. An alternative encoding basis in CV consists of ``Schr\"odinger cat" states \cite{Ourjoumtsev2006,Cultivation,UlanovNOON} $\ket{\Theta_\pm}=\mathcal{N}_\pm(\ket{\gamma}\pm\ket{-\gamma})$, where $\mathcal{N}_\pm=1/\sqrt{2\pm2e^{-2\gamma^2}}$ is the normalization factor.

A missing central piece in the technology of electromagnetic coupling of different physical systems is a method for interconverting between DV and CV encodings of the electromagnetic qubit. Important achievements towards this challenge have been reported in 2014 by two groups \cite{Jeong14, Morin}. They constructed an entangled state between the CV qubit and a ``single-rail" DV qubit in which the logical value is encoded in a single photon being present or absent in a certain mode. Subsequently, this state has been employed as a resource for rudimentary quantum teleportation between these qubits \cite{Ulanov2017}.

However, the single-rail encoding of the qubit is much less common in practical quantum optical information processing than its dual-rail counterpart. This is because single-rail encoding complicates single-qubit operations \cite{Berry2006} and also enhances the qubit measurement errors associated with optical losses and inefficient  detection. Here we address the challenge of conversion between the CV and dual-rail single-photon qubits by preparing a resource of the form 
\begin{align}
\label{eq1}
\ket{R}&=\alpha\ket{H}\ket{\Theta_{+}}+\beta\ket{V}\ket{\Theta_{-}}
%\\&=\dfrac{\ket{1}_H\ket 0_V\ket{\Theta_{+}}+\ket{0}_H\ket1_V\ket{\Theta_{-}}}{\sqrt{2}}
\end{align}
and showing basic applications of it, such as remote state preparation, teleportation and entanglement swapping between the two encodings. %The second line of Eq.~\eqref{eq1} presents the resource state in terms of second quantization, which we will be utilizing from now on. 

\begin{figure*}[t]
	\includegraphics[width=\textwidth]{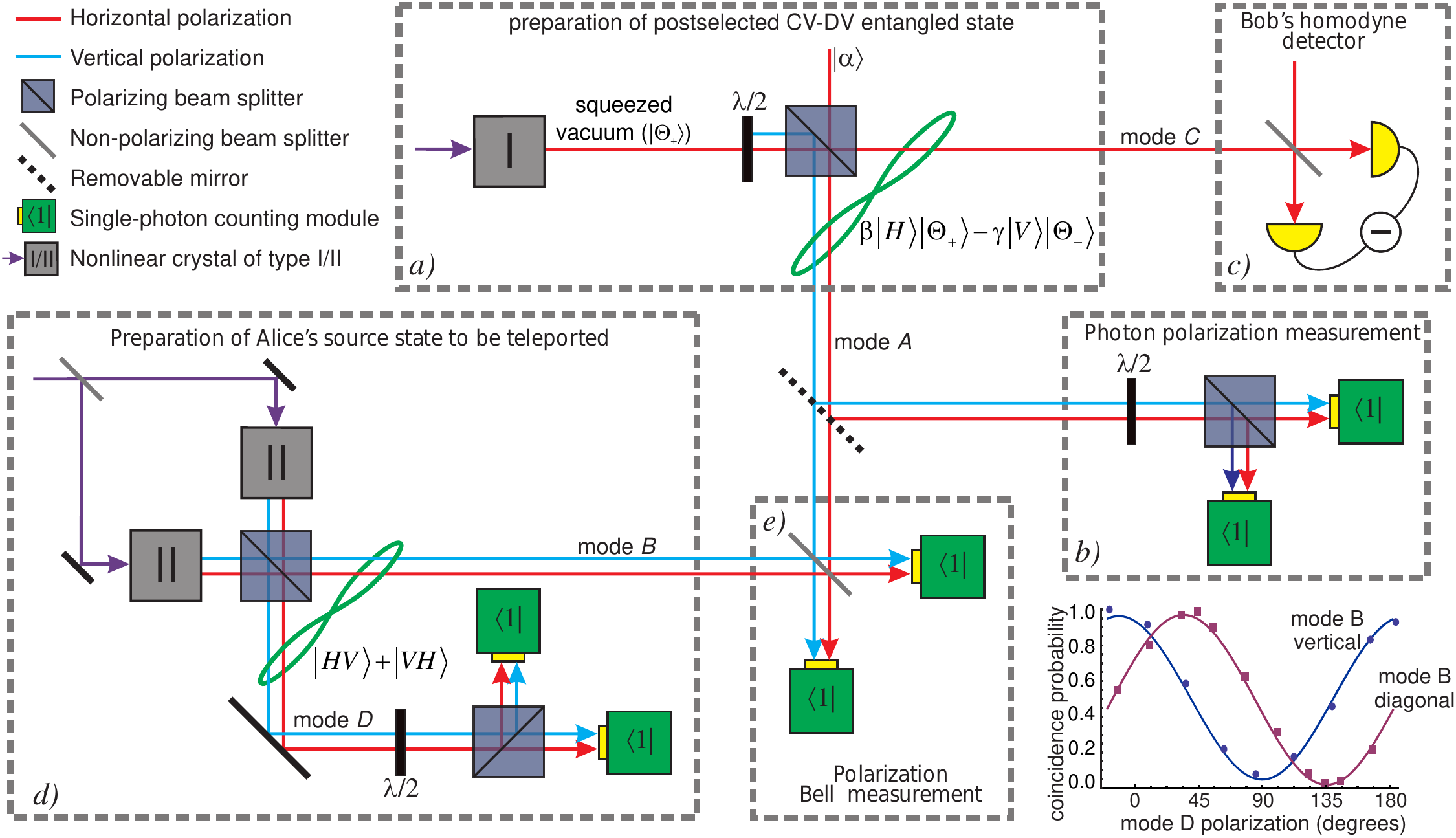}
	\caption{  Conceptual scheme of the experiment. a) State \eqref{eq2} is prepared in modes $A$ and $C$. This state is equivalent to  the discrete-continuous entangled state (\ref{eq1}) conditioned on the presence of a photon in mode $A$. If the photon in mode $A$ comes from $\ket{\Theta_+}$ (i.e.~is horizontally polarized), the state in mode $C$ becomes $\ket{\Theta_-}$. If the photon comes from the vertically polarized state $\ket\alpha$, the state in mode $C$ remains $\ket{\Theta_+}$. The entanglement is verified by measuring the polarization of the photon in the discrete mode (b) and performing homodyne tomography (c) of the state in the continuous mode. d) Preparation of the heralded photon in mode $B$ whose polarization is used as the source state for teleportation. e) Polarization Bell measurement teleports that state onto mode $C$. The mirror leading to part (b) is removed for the teleportation experiment. The inset shows the coincidence rate for simultaneous polarization measurements in modes $B$ and $D$ as a function of the polarization projection angle in mode $D$ while a polarizer is set in mode $B$ to project it onto either horizontal or diagonal polarizations.} 
		%a) , teleportation a DV qubit encoded in H/V basis to a continues variables qubit(\textbf{b}) and for the preparation of the resource state ($\ref{eq1}$) (\textbf{c}).  The full scheme of experimental set-up(\textbf{d}), which corresponds to conceptual scheme (\textbf{c}).  A squeezed vacuum state is generated in the  process of type I degenerate parametric down-conversion (PPKTP). 10\% of the photon flux is mixed with  horizontally polarized weak coherent state in the  spatial mode A, which directed to a BS 50:50, where  it meets with the one of the modes of the polarization-entangled state $\frac{\ket{HV}-\ket{VH}}{\sqrt{2}}$. Simultaneous clicks of  single photon counting modules(SPCM1 and SPCM2) heralds the successful Bell measurement, what corresponds to a generation of the desired entangled state \ref{eq1} in modes C and D. For characterization of  the state of polarization entangled photon pair  the dependence of the probability of a coincidence clicks(\textbf{e}) is measured by means of polarization analyzer(half-wave and quarter-wave plates) and SPD in each output channel(modes B and D). Horizontal axis corresponds to the angle of  one of half-wave plates, while another one set to $\pi/4$ or $3\pi/4$(see Methods): dots represent experiment, solid line - theoretical fit. The visibility calculated from these curves is about 97\%.}
	\label{f1}
\end{figure*}

Developing this entangled resource is also important in the context of purely optical quantum communications. Indeed, the two encodings have complementary advantages \cite{Park}. Continuous variables can benefit from unconditional operations, high detection efficiencies, unambiguous state discrimination and more practical interfacing with conventional information technology. However, they suffer from strong sensitivity to losses and intrinsically limited fidelities. On the other hand, DV approaches can achieve fidelity close to unity, but usually at the expense of probabilistic implementations. Combining the two in hybrid architectures \cite{Loock,HybridQuantum} may offer significant advantages \cite{Near-deterministic,Morin2013}, particularly in the context of 
%quantum key distribution protocols, their security analysis \cite{Rigas,Wittmann} and 
quantum repeaters \cite{Brask2010,QuantumRepeater,Brask2012}

\begin{figure*}[t]
	\includegraphics[width=\textwidth]{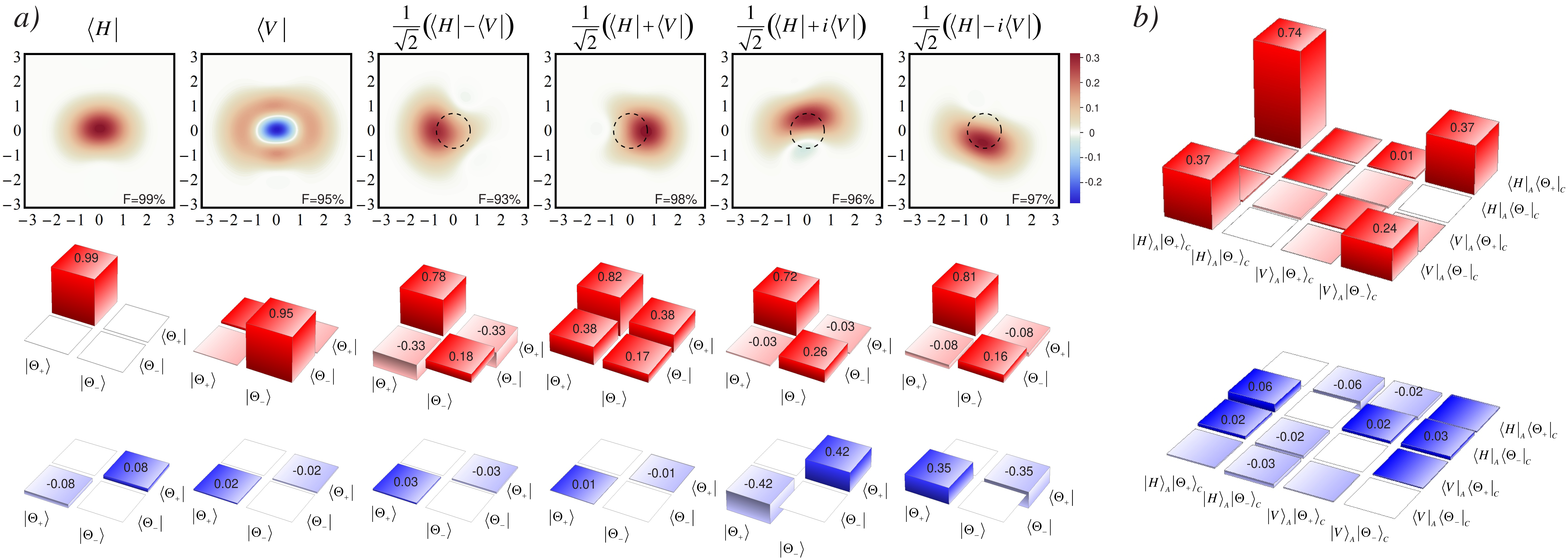}
	\caption{Results of the remote state preparation experiment. a) The state of the the CV mode $C$ when the DV mode $A$ is projected onto various polarization states. Top row: Wigner functions. The fidelities between the experimentally acquired states and theoretically expected superpositions of cat states are shown. Bottom rows: density matrices in the basis $\{\ket{\Theta_+},\ket{\Theta_-}\}$. The real parts are shown in red, imaginary in blue. b) Density matrix of the reconstructed DV-CV state in modes $A$ and $C$.}
	\label{f2}
\end{figure*}

\textbf{Concept.} A simpler version of state  (\ref{eq1}) can be produced as  sketched in Fig.~\ref{f1}(a). We start with a weakly squeezed vacuum state with the squeezing parameter $\zeta = 0.18$, generated in a horizontally polarized mode via degenerate parametric down-conversion. This state is an excellent approximation to the low-amplitude positive Schr\"odinger's cat state $\ket{\Theta_{+}}$  with the  amplitude $\gamma_{+}=\sqrt{\zeta}$ \cite{Ourjoumtsev2006,Ulanov2017}. It then passes through a half-wave plate and polarizing beam splitter (PBS), which in combination act as a variable-reflectivity beam splitter. In the case of low reflectivity $r\ll1$,  the resulting state can be written as 
\begin{align*}
\ket\psi_{VA,HC}=&\ket{0}_{VA}\otimes\ket{\Theta_{+}}_{HC}+\sqrt{r}\hat a^\dag\ket{0}_{VA}\otimes\hat a\ket{\Theta_+}_{HC}\\
=&\ket{0}_{VA}\otimes\ket{\Theta_{+}}_{HC}+\sqrt{r} \ket{1}_{VA}\otimes\frac{\mathcal{N}_+}{\mathcal{N}_-}\ket{\Theta_{-}}_{HC},
\end{align*} because applying a photon annihilation operator to the state $\ket{\Theta_{+}}$ transforms it into $\ket{\Theta_{-}}$ \cite{Ourjoumtsev2006}. In the above equation, $VA$ and $HC$ denote, respectively, the vertical component of spatial mode $A$ and the horizontal component of mode $C$.

Now let us suppose a weak horizontally polarized coherent state $\ket{\alpha}\approx\ket{0}+\alpha\ket{1}$ is injected into the input mode $A$ of the PBS. We then obtain the state  
\begin{align}\label{eq2}
\ket{\Omega}_{AC}&= \ket\alpha_{HA}\ket\psi_{VA,HC}\approx \ket{0}_{HA}\ket{0}_{VA}\ket{\Theta_{+}}_{HC}\\
& +\alpha\ket{1}_{HA}\ket{0}_{VA}\ket{\Theta_{+}}_{HC}+\beta\ket{0}_{HA}\ket{1}_{VA}\ket{\Theta_{-}}_{HC}\nonumber ,\nonumber
\end{align}
where $\beta=\sqrt{r}{\mathcal{N}_+}/{\mathcal{N}_-}$ and we approximated to the first order in $\beta$ and $\alpha$. The last line in the above state corresponds to a single photon present in spatial mode $A$. It comprises the desired resource \eqref{eq1} in modes $A$ and $HC$ (we use $\ket V_A\equiv \ket{0}_{HA}\ket{1}_{VA}$ and $\ket H_A\equiv\ket{1}_{HA}\ket{0}_{VA}$ to switch between the first and second quantization notations). 

In spite of a strong vacuum component in mode $A$, state  \eqref{eq2} can be utilized  for quantum communication protocols. Here we demonstrate the application of this state for remote state preparation \cite{RemoteStatePrep} and teleportation \cite{Bennett1993} of a qubit from the polarization onto the CV encoding. Moreover, we show how to utilize entanglement swapping to purge the vacuum component from that state.

%This can be verified experimentally by measuring the photon in mode $A$ in various polarization bases and 

For remote state preparation, we project mode $A$ onto a superposition $a\ket{H}+b\ket{V}$ by means of a polarization analyzer and a single-photon detector (SPCM) [Fig.~\ref{f1}(b)]. A click of the SPCM heralds the preparation of a CV qubit 
\begin{equation}\label{RSP}
(a^*\bra{H}+b^*\bra{V})_A\ket\Omega_{AC}=a^*\alpha\ket{\Theta_{+}}_C+b^*\beta\ket{\Theta_{-}}_C
\end{equation}
(we use letter $C$ to denote mode $HC$ from now on, because mode $VC$ is in the vacuum state and does not become involved in the analysis). The performance of the procedure is tested by homodyne tomography in mode $C$ [Fig.~\ref{f1}(c)]. The values of $a$ and $b$ can be varied arbitrarily by changing the angles of half- and quarter-wave plates in the polarization analyzer; thereby CV qubits of arbitrary values can be conditionally prepared~\cite{Nielsen2010}. 

For teleportation, ``Alice" prepares a heralded source photon in  a polarization state $\ket\chi=a\ket H+b\ket V$  in an  additional spatial mode $B$ [Fig.~\ref{f1}(d)]. We then apply the  Bell state projector $\bra{\Psi^-}=(\bra{H}\bra{V}-\bra{V}\bra{H})/\sqrt{2}$ to  modes $A$ and $B$ as shown in Fig.~\ref{f1}(e). We obtain
\begin{equation}
\label{eq3}
\begin{aligned}
\ket{\varphi}_C=&\bra{\Psi^-}_{AB}(\ket{\Omega}_{AC}\ket{\chi}_{B})\\=& \frac1{\sqrt2}\left(a\beta\ket{\Theta_{-}}_{C}-b\alpha\ket{\Theta_{+}}_{C}\right). 
\end{aligned}
\end{equation}
Because the Bell measurement requires two photons to be present in its input, it will cut off the vacuum term in Eq.~(\ref{eq2}), so the state of the input photon is teleported onto ``Bob's" freely propagating CV qubit in mode $C$. 

%Full realization of this scheme, with complete elimination of the vacuum term, would require a heralded polarization-entangled photon pair source. Such sources are  accessible with existing experimental methods \cite{heraldedHVVH1,heraldedHVVH2}. However, using this technique in our experiment would result in an extremely low state production rate so we would not be able to characterize it. Therefore, for our proof-of-principle demonstration, we use a standard postselected entangled pair source that has been routinely utilized in quantum optics labs since 1995 \cite{Kwiat95}. 

\textbf{Experiment: remote state preparation.}  We prepare state (\ref{eq2}) as described above (see Methods for further detail), with the ratio $\beta/\alpha$ of about $0.6$. In order to create a maximally entangled DV-CV state, this ratio should have been unity. However, choosing a lower value helps increasing the data acquisition rate (which was a  critical parameter in this experiment) while still allowing us to see the effects we wish to observe.   %The initial single-mode squeezed vacuum state is prepared with the squeezing parameter $\zeta = 0.18$, which approximates  $\ket{\Theta_{+}}$  with the  amplitude $\gamma_{+}=\sqrt{\zeta}$\cite{Ulanov2017}.  The half-wave plate, in combination with the PBS, has a reflectivity of  $r^2=0.1$. % n we send SMSV to a system of half-wave plate and PBS, which reflects 10\% of photon flux. To achieve the desired state, the phase difference between the tho PBS inputs must be locked (see Methods).  

We project mode $A$ onto elements of the canonical, diagonal and circular polarization bases and  perform homodyne tomography on the resulting states in mode $C$. A total of 2500 quadrature samples are recorded for each state.  The states are then reconstructed via a maximum-likelihood algorithm \cite{Lvo2004} with a correction for the homodyne detection efficiency of 0.55 [Fig.~\ref{f2}(a)]. 

Projections onto the horizontal and vertical polarization states yield $\ket{\Theta_{+}}$ and $\ket{\Theta_{-}}$, respectively,  which resemble the squeezed vacuum and squeezed single-photon states \cite{Ourjoumtsev2006}. Our results show  fidelities of 0.99 and 0.95 with the ideal cat states of amplitudes  $\gamma_+=0.45$ and $\gamma_-=0.90$, respectively, where the fidelity  between states $\hat\rho_1$ and $\hat\rho_2$ is defined as $F=\left(\tr\sqrt{\hat\rho_1^\frac12\hat\rho_2\hat\rho_1^\frac12}\right)^2$. A difference between $\gamma_+$ and $\gamma_-$ is inherent in the preparation method; theoretically, we expect $\gamma_-/\gamma_{+}=\sqrt{3}$ \cite{Ulanov2017}. Note that the Wigner function of the experimentally reconstructed state $\ket{\Theta_{-}}$ exhibits negativity even without efficiency correction. 

Projecting mode $A$ onto superpositions of $\ket H$ and $\ket V$ produces analogous CV superpositions \eqref{RSP}.  In particular, projecting onto the diagonal basis  yields coherent states whose Wigner functions have a characteristic shape of a displaced Gaussian peak. The coherent nature of these superpositions is also evidenced by their density matrices in the basis $\{\ket{\Theta_+},\ket{\Theta_-}\}$ shown in Fig.~\ref{f2}(b). The fidelity with the theoretically expected states  \eqref{RSP} exceeds 0.93  in all cases.  %Fidelity between reconstructed states and corresponding superposition  $\ket{\Theta_{-}}\pm e^{i\phi}\ket{\Theta_{+}}$ in each case is about 95\%. 

The states displayed in Fig.~\ref{f2}(a) can be used to fully reconstruct the component of the DV-CV state in modes $A$ and $C$, projected onto the subspace corresponding to a single photon in mode $A$ (see Methods). The density matrix of this state [Fig.~\ref{f2}(c)] has a fidelity of 0.93 with the state \eqref{eq1}. The fidelity with the maximally entangled state, given by \eeqref{eq1} with $\alpha=\beta$, equals 0.84.  This evidences entanglement of this state, because the fidelity of a separable state with a maximally entangled biqubit state cannot exceed $1/2$ \cite{Sackett2000}.

%the lower bound of the fidelity between  the discrete-continuous state in modes $A$ and $C$, conditioned on the photon detection in mode $A$ and state \eqref{eq1}, as $0.85\pm0.02$ \cite{Blinov}.

\begin{figure*}[t]
	\includegraphics[width=\textwidth]{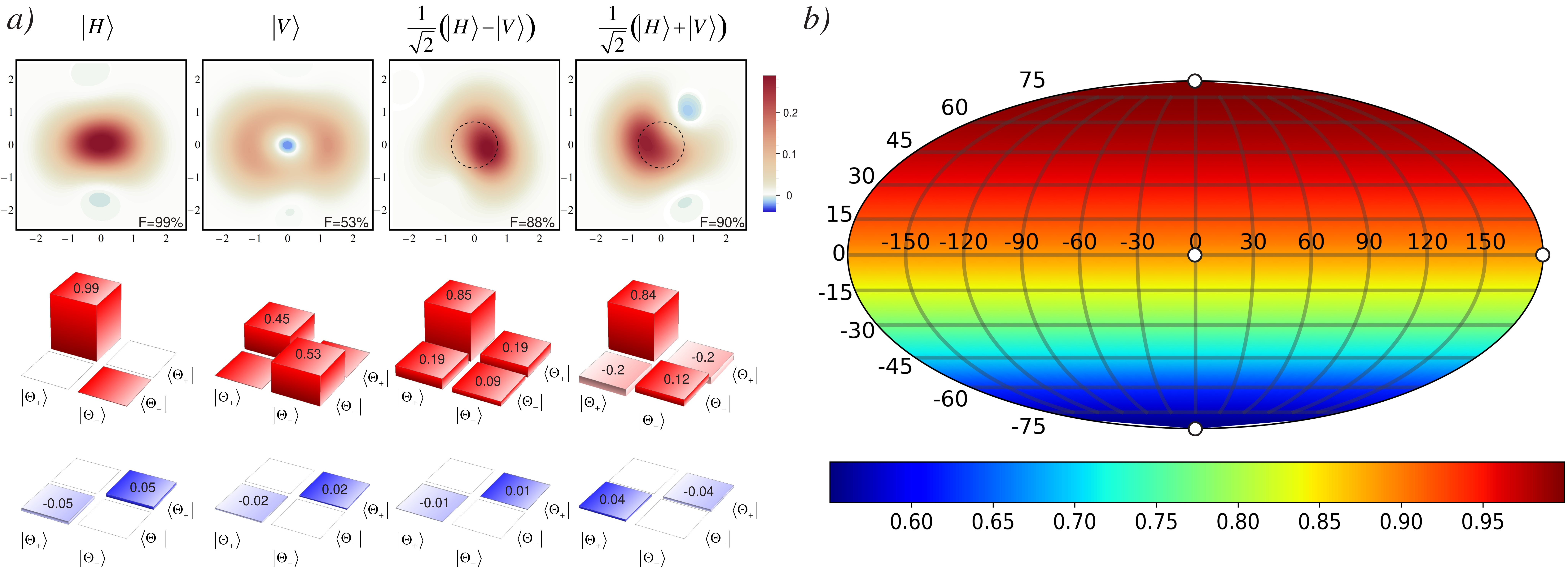}
	\caption{Results of the teleportation from a polarization qubit (mode $B$) onto a CV qubit (mode $C$). a) Wigner functions and density matrices of the teleported states for four input polarization states [same notation as in Fig.~2, the fidelities are calculated with respect to the theoretically expected superpositions \eqref{eq3} of cat states]. b) Teleportation fidelity based on the theoretical model \eqref{QTtheory} which agrees well with the experimental results in (a), calculated for the entire Bloch sphere. The white dots represent the  input states of the teleportation experiment.}
	\label{f3}
\end{figure*}

\textbf{Experiment: teleportation.} To produce a source photon, we first prepare a photon pair in a polarization entangled Bell state $\ket{\Psi^+}_{BD}$. This is realized by overlapping the outputs of parametric down-conversion from two crystals, in each of which a collinear, frequency-degenerate pair of the form $\ket{HV}$ is produced, on a polarizing beam splitter [see Fig.~\ref{f1}(d), Ref.~\cite{HVVHsourceGisin} and Methods]. By measuring the polarization of the photon in mode $D$, we prepare a heralded photon in mode $B$ in a certain polarization state. The performance of the method is illustrated by the inset in Fig.~\ref{f1}.

Subsequently, we perform a Bell measurement on modes $A$ and $B$ using the technique from the original experiment on quantum teleportation of a photon polarization state \cite{teleportZeilinger}. Namely, these modes are subjected to interference on a symmetric non-polarizing beam splitter. If a single photon is present in each beam splitter input, the Hong-Ou-Mandel effect \cite{HOM} forces the photons to emerge in the same output spatial mode unless the input is in the Bell state $\ket{\Psi^-}_{AB}$ [Fig.~\ref{f1}(e)]. Thus a coincidence detection event in both beam splitter outputs projects its input onto that state. 

This method however suffers from an issue raised in Braunstein and Kimble's correspondence \cite{Aposterioriteleportation} on Ref.~\cite{teleportZeilinger}. The coincidence event can occur not only due to one photon coming from each of the modes $A$ and $B$, but also when both photons come from the same mode. The latter events, which we refer to as ``double $A/B$ events" result in false positive Bell state detection. In our case, only the double $B$ events are of concern because the photon in mode $B$ is heralded and $\alpha,\beta\ll 1$. The probability $p_{\rm dB}$ of these events can be reduced by lowering the pumping of parametric down-conversion in modes $B$ and $D$ \cite{Pan2003} to the level such that $p_{\rm dB}\ll p_{\rm good}$, where $p_{\rm good}$ is the probability of a true positive Bell detection event in which the two photons come from different modes. In the actual experiment, we have $p_{\rm good}/p_{\rm dB}=1$--3 (see Methods). The variation of this ratio is due to the dependence of the ``good" Bell detection probability on the input state: $p_{\rm good}\propto\|\ket\varphi\|^2=(|a\beta|^2+|b\alpha^2|)/2$,  where the state $\ket\varphi$ is given by Eq.~\eqref{eq3}. %By lowering, we set $p_B\approx 0.25 p_A$. %This is a compromise between the above requirement and the need for an photon production rate that would permit collecting enough data for homodyne tomography within a reasonable time.At the same time, the rate of correct Bell detection events is proportional to the probability . 

We implement the teleportation for the source photon states $\ket{H}$, $\ket{V}$ and  $(\ket{H}\pm\ket{V})/\sqrt 2$. For each of these  states, we perform homodyne tomography of the state in the output CV mode $C$ conditioned on a triple coincidence photon detection in modes $A$, $B$ (Bell detection) and one of the photons in mode $D$ (polarization state heralding).  The rate of these events is about 0.015 Hz. For each input state, about 1500 quadrature samples are collected.

Figure \ref{f3}(a) shows the states reconstructed from the results of this measurement, with compensation for the homodyne detection efficiency of $0.55$. The primary detrimental effect on the teleportation fidelity is false positive Bell state detections due to double $B$ events. In such an event, no photon annihilation in mode $C$ takes place, resulting in that mode containing the state $\ket{\Theta_{+}}$. As a result, the teleportation output can be written as
\begin{equation}\label{QTtheory} \hat\rho_{\rm out}=\frac{p_{\rm good}}{p_{\rm good}+p_{\rm dB}}\mathcal{N}(\ketbra{\varphi}{\varphi})+\frac{p_{\rm dB}}{p_{\rm good}+p_{\rm dB}}\ketbra{\Theta_{+}}{\Theta_{+}}, 
\end{equation} where $\mathcal{N}$ denotes normalization. The ratio $p_{\rm dB}/p_{\rm good}$ reaches a maximum value of $\sim 1$ for the input state $\ket\chi=\ket H$ (so $a=1$ and $b=0$), in which case the fidelity of the output state approaches 53\% [Fig.~3(a), second column]. For input states with a significant fraction of $\ket V$, the fidelity is much higher.

The simple model of \eeqref{QTtheory} describes the four teleported states with a $\ge$93\% fidelity. We use the above model to determine the teleportation fidelity for an arbitrary polarization qubit as input [Fig.~3(b)]. We find the mean fidelity over the full Bloch sphere to equal 80\%, which is above the classical benchmark of $2/3$ \cite{Massar1995}.

%The obtained states are of smaller amplitudes and have lower fidelities with the theoretically expected states than in the experiment without entanglement swapping. These differences  are caused mainly by two reasons. First, a non-ideal mode overlap of modes $A$ and $B$ compromises the Bell measurement. Second, the finite ratio $p_B/p_A$ leads to a significant fraction of false positive Bell detection events due to both photons  originating from mode $B$. Nevertheless, by analyzing the four reconstructed states in the same manner as discussed previously, we recover a fidelity of ??? with the maximally entangled state, which is above the  upper bound for separable states.  

\textbf{Entanglement swapping.} An important alternative interpretation of our experiment is entanglement swapping \cite{Pan1998}. If one considers the Bell measurement in modes $A$ and $B$ without accounting for the measurement in mode $D$, one obtains
\begin{equation}
\label{EntSwap}
\begin{aligned}
&\bra{\Psi^-}_{AB}(\ket{\Omega}_{AC}\ket{\Psi^+}_{BD}) \\
&\hspace{2cm}=\alpha\ket{H}_{D}\ket{\Theta_{+}}_{C} - \beta\ket{V}_{D}\ket{\Theta_{-}}_{C}. 
\end{aligned}
\end{equation}
The Bell measurement cuts off the first term in Eq.~(\ref{eq2}), thereby heralding a freely propagating resource state (\ref{eq1}) in modes $C$ and $D$.

A proper realization of this scheme would require a heralded source of photon pairs in a Bell state. Such sources are  accessible with existing experimental methods \cite{heraldedHVVH1,heraldedHVVH2,heraldedHVVH3}. However, constructing such a source in our experiment would result in a prohibitively low state production rate. A ``conventional" Bell state source with a dominating vacuum component, such as ours, will result in a large fraction of false positive double $A$ or double $B$ Bell state detection events. This will preclude the production of state (\ref{EntSwap}) with high fidelity.

However, the results in Fig.~\ref{f3}(a) can be interpreted to characterize the entanglement swapping output state in the \emph{a posteriori} manner: conditioned on a photon detection event in mode $D$. This event guarantees that a photon has been present in mode $B$ and thereby eliminates the double $A$ events from the analysis. 

The four reconstructed states can be used to determine a lower bound of the fidelity between the entanglement swapping output and the maximally entangled state \cite{Blinov} (see Methods). We find this lower bound to be 0.56$\pm$0.02, which is above the entanglement threshold of $1/2$ \cite{Sackett2000}.
As a concluding remark, the technique reported here can be useful for the interconversion not only between polarization and continuous-variable qubits, but also between single-rail and dual-rail photonic qubits. This is because, in the limit of low squeezing, the state $\ket{\Theta_+}$ is close to the vacuum state while the state $\ket{\Theta_-}$ is well-approximated by the single photon. However, the specific scheme of our experiment may not be optimal for this application because the production rate of the corresponding entangled states would tend to zero in this limit. In the future, we plan to study possible modifications of the scheme that would result in higher productivities.

\acknowledgments{We thank Marco Barbieri for inspiring discussions. AL's research is supported by NSERC and CIFAR.}

\section{Methods} \paragraph{Experimental setup.} The initial  1.5-dB  single-mode squeezed vacuum state in mode $C$ is produced by degenerate parametric down-conversion in a  periodically poled potassium titanyl phosphate crystal (PPKTP, Raicol) under type-I phase-matching conditions. The crystal is pumped with $\sim$20 mW frequency-doubled radiation of the master laser (Ti:Sapphire Coherent Mira 900D, with a wavelength of 780 nm,
repetition rate of $R=76$ MHz and pulse width of ∼1.5 ps) \cite{Huisman}. 

For the preparation of polarization-entangled photon pairs [Fig.~1(d)], a polarization interferometer scheme is used \cite{HVVHsourceGisin}. A symmetric beam splitter at the entrance of the interferometer splits the $\sim$5-mW pump beam into two equal parts that are directed into PPKTP crystals in each path. Parametric down conversion in each crystal occurs in a collinear, frequency-degenerate, type-II regime and generates a two-mode squeezed vacuum state $\ket{0}_H\ket{0}_V+\lambda\ket{1}_H\ket{1}_V+\lambda^2\ket{2}_H\ket{2}_V+O(\lambda^3)$. After a polarizing beamsplitter at the end of interferometer, the orthogonally polarized modes from the two arms become temporarily and spatially indistinguishable in each of the two output modes B and D.

The path length difference between the two arms of the interferometer is locked. The feedback for the lock is obtained from the interferometric signal of two pump beams and is applied to a piezoelectric transducer  in one of the arms. The resulting polarization-entangled state is characterized by simultaneous polarization analysis in modes $B$ and $D$ in linear bases.  %State characterization is made by means polarization analyzers(half-wave plate and quarter-wave plate) and single-photon detectors in each output mode. 
The coincidence count rate is measured with different angles of the half-wave plate in mode D while a polarizer inserted in front of the detector in mode $B$ is kept constant. This rate exhibits a characteristic sinusoidal shape with a visibility  of 97\% (Fig.~\ref{f1}, inset).

Photon detection is implemented by fiber-coupled SPCMs (Excelitas). In mode $D$, two SPCMs are used to detect orthogonal polarization states, which permits simultaneous acquisition of the quadrature data corresponding to the teleportation of both these states. The data acquisition is triggered by a home-made delay/coincidence circuit based on  Artix-7 35T field-programmable gate array.

The relative phase between the two terms of the CV-DV entangled state is determined by the phases of the input coherent and squeezed states, whose difference must therefore be kept constant. We measure both phases with respect to that of the local oscillator used for homodyne detection.  The phase of the squeezed vacuum is determined from the quadrature variance acquired by the homodyne detector in $HC$ without conditioning on single-photon detection events. To measure the phase of the coherent state, we prepare it with a significant vertical polarization component. This component is then reflected into the mode $VC$ and measured with an auxilliary homodyne detector. The difference between the two phases is locked to zero by means of a feedback signal applied to a piezoelectric transducer in  the path of the input coherent state.

%Outputs  in a Bell measurement apparatus of the beamsplitter(ratio of reflected power to transmitted through beamsplitter is 4:3) are directed  to   Coincident click of three SPCMs( two in Bell projection and one  in discrete mode analyzer) heralds the preparation of state (\ref{eq1}) in continues mode C.  The preparation rate of the state in mode C is about $\sim$0.03Hz.

%For each of four states about 3000 points are acquired. For cases with mode D projected on diagonal states $\ket{H}\pm\ket{V}$ only half of the events satisfy the phase conditions, resulting to about 1500 points  for each state. The reconstruction of the states is performed using the iterative maximum-likelihood algorithm \cite{Lvo2004}. The local oscillator phase is varied  by PZT3 and its time-dependent value is extracted from the variance of the non-triggered quadrature data, which corresponds to the single-mode squeezed state. 

The total quantum efficiency of homodyne detection, 55\%, is determined from the analysis of the negative cat state $\ket{\Theta_-}$ generated in mode $C$ conditioned on a photon detection in $VA$. The main efficiency reduction factors are optical losses (90\% cumulative transmissivity of all optical elements, in addition to the tapping beamsplitter which also has a $1-r=90$\% transmissivity), mode matching between the signal and local oscillator (81\%) and the quantum efficiency of the homodyne detector (86\%) \cite{Kumar2012,Masalov2017}.

%In this case, to generate a coherent state of amplitude $\gamma_{+}$, taking into account normalizing factors  $N_{\pm}=1/\sqrt{2\pm 2e^{-2 \gamma_{\pm}^{2}}}$ for Schr\"odinger's cat states $\ket{\Theta_{\pm}} = N_{\pm} (\ket{\gamma_\pm} \pm \ket{-\gamma_\pm})$, we make  the ratio of count rates from coherent state and reflected part of SMSV  equal to $(N_{+}/N_{-})^{2}=(1-e^{-2\gamma_{-}^{2}})/(1+e^{-2\gamma_{+}^{2}})\approx3$. 

\paragraph{Photon count rates.} With mode $A$ blocked, the polarization-entangled pairs generated in the two type-II crystals produce count rates of  $R_{B,D}\sim4\times10^{3}$ s$^{-1}$ in each of the SPCMs in mode $D$ and the Bell detector. The coincidence rate between each pair of SPCMs in mode $D$ and the Bell detector is $\sim20$ s$^{-1}$, meaning that the single-photon detection efficiency is $\eta_{\rm SPCM}=0.01$. Such a low efficiency is explained by the presence of narrowband (0.2 nm) filters in front of each SPCM \cite{Huisman}, in addition to the usual linear losses. Based on these numbers, the probability of an undesired double-B event, coincident with a click in and one of the detectors in mode $D$, can be estimated as $p_{\rm dB}=\frac32\eta_{\rm SPCM}R_B^2/R^2\approx4\times10^{-11}$. 

When mode $B$ is blocked, the count rate in each of the two SPCMs in the Bell detector is $R_\alpha=18\times10^{3}$ s$^{-1}$ due to the coherent state and $R_\beta=6\times10^{3}$ s$^{-1}$ due to the squeezed vacuum state, so $R_\beta/R_\alpha=\beta^2/\alpha^2=1/3$. The probability of a ``good" triple coincidence event of the two SPCMs in the Bell detector and one of the detectors in mode $D$ is therefore estimated as $p_{\rm good}=\eta_{\rm SPCM}R_B(b^2R_\alpha+a^2R_\beta)/R^2\approx4$--$12\times10^{-11}$. The expected total triple coincidence event rate, $R(p_{\rm good}+p_{\rm dB})=6$--$12\times 10^{-3}$ s$^{-1}$, is consistent with the rate observed in the experiment within a factor of one and a half. 

 % corresponding to $p_B=R_B/R=2.6\times10^{-5}$.

\paragraph{Two-mode state reconstruction.}
The DV-CV state $\hat\rho$ in modes $A$ and $C$ [Fig.~2(b)] can be recovered from the six density matrices in mode $C$ [Fig.~2(a)] that represent the projections $\braketop\pi{\hat\rho}\pi$ of that state onto various polarization states $\ket\pi$ in mode $A$. To find $\hat\rho$, we write it in a generic form
\begin{eqnarray*}
	\hat\rho=\ket{H}\bra{H}\otimes\hat\rho_{HH}+\ket{V}\bra{V}\otimes\hat\rho_{VV}\\
	+\ket{H}\bra{V}\otimes\hat\rho_{HV}+\ket{V}\bra{H}\otimes\hat\rho_{HV}^\dagger
\end{eqnarray*}
where, e.g.,~$\hat\rho_{HV}=\bra{H}\hat\rho\ket{V}$. The first two terms in the above expression are obtained directly from the first two columns in Fig.~2(a). The remaining two terms are evaluated from the remaining four columns according to
\begin{eqnarray*}
	\hat\rho_{HV} = \frac{1}{2}\left( \bra{D}\hat\rho\ket{D} - \bra{A}\hat\rho\ket{A} + i\bra{L}\hat\rho\ket{L} - i \bra{R}\hat\rho\ket{R} \right),
\end{eqnarray*} where 
$\ket{A,D} = \frac{\ket{H}\pm\ket{V}}{\sqrt{2}}$ and $\ket{R,L} = \frac{\ket{H}\pm i\ket{V}}{\sqrt{2}}$ are the diagonal and circular polarization states.

\paragraph{Entanglement criterion.}
In contrast to the state of modes $A$ and $C$ analyzed above, the CV-DV state of modes $D$ and $C$, obtained after entanglement swapping, cannot be reconstructed because only its projections onto the canonical and diagonal polarization states are known [Fig.~3(a)]. However, these data can be used to estimate the lower bound of the fidelity with the maximally entangled state $\ket{\Psi_{ME}}=\frac1{\sqrt2}\left(\ket{H}\ket{\Theta_+} -\ket{V}\ket{\Theta_-}\right)$. To this end, we follow the argument of Ref.~\cite{Blinov} and write %Here we reproduce this argument in application to our case. The key feature of the resource state (\ref{eq1}) is its entanglement. In order to prove that our experimental state possesses such property we need to find some criterion. Usual criteria such as negativity don't suit for the entanglement swapping experiment. Because it demands full information about two mode state, i. e., two mode density matrix. We made projection measurements in two discrete bases. As explained earlier the full reconstruction requires measurements in three bases.

%But information about two projections sets on two bases allows to compare experimental state with the maximally entangled state  in terms of fidelity:

\begin{align}\label{FBlin1}
	&F = \bra{\Psi_{ME}} \hat\rho \ket{\Psi_{ME}} \\
	 &=\frac12\big( \hat\rho_{H\Theta_+, H\Theta_+} +  \hat\rho_{V\Theta_-, V\Theta_-} - \hat\rho_{H\Theta_+, V\Theta_-} -  \hat\rho_{V\Theta_-, H\Theta_+} \big)\nonumber
\end{align}
where $\hat\rho$ is the density matrix of the DV-CV state in question written in the basis $\{\ket H,\ket V\}\otimes\{\ket{\Theta_+},\ket{\Theta_-}\}$. The first two terms in \eeqref{FBlin1} are obtained from the first two columns in Fig.~3(a) taking into account the probabilities of occurrence of the corresponding polarization states in mode $D$. The sum of the last two terms can be estimated as follows:
\begin{align}\label{FBlin2}
&\hat\rho_{H\Theta_+,V\Theta_-} + \hat\rho_{V\Theta_-, H\Theta_+}  \\
&=\hat\rho_{D\Theta_D, D\Theta_D} + \hat\rho_{A\Theta_A, A\Theta_A} - \hat\rho_{D\Theta_A, D\Theta_A} - \hat\rho_{A\Theta_D, A\Theta_D}\nonumber \\&-
	(\hat\rho_{V\Theta_+, H\Theta_-} + \hat\rho_{H\Theta_-, V\Theta_+}) \nonumber
\end{align}
where $\ket{\Theta_{A,D}}=\frac1{\sqrt2}(\ket{\Theta_+}\pm\ket{\Theta_-})$. The first line in the right-hand side of the above equation is obtained from the last two columns in Fig.~3(a). The last line can be bounded by 
\begin{equation}\label{FBlin3}|\hat\rho_{V\Theta_+, H\Theta_-} + \hat\rho_{H\Theta_-, V\Theta_+}|\le
2\sqrt{\hat\rho_{H\Theta_-, H\Theta_-} \hat\rho_{V\Theta_+, V\Theta_+}}.
\end{equation}     
Combining Eqs.~\eqref{FBlin1}, \eqref{FBlin2} and \eqref{FBlin3} yields a bound on the fidelity $F$. Since the lower bound exceeds $1/2$, the state is entangled \cite{Sackett}.


\begin{thebibliography}{99}



\bibitem{Kur15} Kurizki, G., Bertet, P., Kubo, Y., M\o{}lmer, K., Petrosyan, D., Rabl, P. \& Schmiedmayer. J. Quantum technologies with hybrid systems. \textit{Proc. Natl. Acad. Sci.} \textbf{112}, 3866--3873 (2015). 

\bibitem{Kok} Kok, P. et al. Linear optical quantum computing with photonic qubits. \textit{ Rev. Mod. Phys.} \textbf{79}, 135–174 (2007).

\bibitem{Braunstein} Braunstein, S. L. \& Pati, A. (eds) Continuous Variable Quantum Information (Kluwer Academic, 2003).

\bibitem{Jeong} Jeong, H. \& Kim, M. S. Efficient quantum computation using coherent states. \textit{ Phys. Rev. A} \textbf{65}, 042305 (2002).

\bibitem{Ralph2003} Ralph, T. C., Gilchrist, A.,  Milburn, G. J.,  Munro, W. J. \& Glancy, S. Quantum computation with optical coherent states. \textit{Phys. Rev. A} \textbf{68}, 042319 (2003).

\bibitem{Ourjoumtsev2006} Ourjoumtsev, A., Tualle-Brouri, R., Laurat, J. \&  Grangier, P. Generating optical schrodinger kittens for quantum information processing. \textit{Science} \textbf{312}, 83--86 (2006).

\bibitem{Cultivation} Sychev, D. V., Ulanov, A. E., Pushkina, A. A., Richards, M. W., Fedorov, I. A., and Lvovsky, A. I. Enlargement of optical Schr\"odinger's cat states. \textit{Nat. Photon.} \textbf{11}, 379–382 (2017).

\bibitem{UlanovNOON} Ulanov, A. E., Fedorov, I. A., Sychev, D., Grangier, P. \& Lvovsky, A. I. Loss-tolerant state engineering for quantum-enhanced metrology via the reverse
Hong-Ou-Mandel effect. \textit{Nat. Commun.} \textbf{7}, 11925 (2016).

\bibitem{Morin} Morin, O., Huang, K., Liu, J., Le Jeannic, H., Fabre, C. \&  Laurat, J.  Remote creation of hybrid entanglement between particle-like and wave-like optical qubits. \textit{Nat. Photon.} \textbf{8}, 570–574 (2014).

\bibitem{Jeong14} Jeong, H., Zavatta, A., Kang, M., Lee, S., Costanzo, L. S., Grandi, S. Ralph, T. \&  Bellini, M. Generation of hybrid entanglement of light. \textit{Nat. Photon.} \textbf{8}, 564–-569 (2014).

\bibitem{Ulanov2017} A. E. Ulanov, D. Sychev, A. A. Pushkina, I. A. Fedorov and A. I. Lvovsky, A scheme for quantum teleportation between discrete and continuous encodings of an optical qubit. \textit{Phys. Rev. Lett.} {\bf 118}, 160501 (2017).

\bibitem{Berry2006} D. W. Berry, A. I. Lvovsky, B. C. Sanders, Interconvertibility of single-rail optical qubits. \textit{Opt. Lett.} {\bf 31}, 107 (2006).

\bibitem{Park} Park, K. \& Jeong, H. Entangled coherent states versus entangled photon pairs for practical quantum-information processing.  \textit{Phys. Rev. A} \textbf{82}, 062325 (2010).

\bibitem{Loock} Van Loock, P. Optical hybrid approaches to quantum information. \textit{ Laser Photon. Rev.} \textbf{5}, 167–200 (2011).

\bibitem{HybridQuantum} Andersen, U. L., Neergaard-Nielsen, J. S., van Loock, P.  \& Furusawa, A. Hybrid quantum information processing. \textit{Nat. Phys.} \textbf{11}, 713–719(2015).

\bibitem{Morin2013}  Morin O. et al. Witnessing trustworthy single-photon entanglement with local homodyne measurements. \textit{ Phys. Rev. Lett.} \textbf{110}, 130401 (2013).

\bibitem{Near-deterministic} Lee, S.-W. \& Jeong, J. Near-deterministic quantum teleportation and resource-eﬃcient quantum computation using linear optics and hybrid qubits. \textit{Phys. Rev. Lett.} \textbf{87}, 022326 (2012).

\bibitem{Brask2012} J. Borregaard, J. B. Brask, and A. S. Sørensen, Hybrid quantum repeater protocol with fast local processing. \textit{Phys. Rev. A} {\bf 86}, 012330

\bibitem{QuantumRepeater} Sangouard, N. et al. Quantum repeaters with entangled coherent states. \textit{J. Opt. Soc. Am. B} \textbf{27}, A137-A145 (2010).

\bibitem{Brask2010}  J. B. Brask, I. Rigas, E. S. Polzik, U. L. Andersen, and A. S. S\o{}rensen. Hybrid Long-Distance Entanglement Distribution Protocol. \textit{Phys. Rev. Lett.} {\bf 105}, 160501 (2010).

\bibitem{RemoteStatePrep} Bennett, C. H. {\it et al.}   Remote State Preparation. \textit{Phys. Rev. Lett.} \textbf{87}, 077902 (2001).

\bibitem{Bennett1993} Bennett, C. H., Brassard, G., Crépeau, C., Jozsa, R., Peres, A. \&  Wootters, W. K. Teleporting an unknown quantum state via dual classical and Einstein-Podolsky-Rosen channels. \textit{Phys. Rev. Lett.} \textbf{70}, 1895 (1993)

\bibitem{Nielsen2010} Neergaard-Nielsen, J. S., Takeuchi, M., Wakui, K., Takahashi, H., Hayasaka, K., Takeoka, M. \&  Sasaki, M. Optical Continuous-Variable Qubit. \textit{Phys. Rev. Lett.} \textbf{105}, 053602(2010)

\bibitem{Lvo2004}  Lvovsky, A. I. Iterative maximum-likelihood reconstruction in quantum homodyne tomography. \textit{J. Opt. B.} \textbf{6}, S556--S559 (2004).

\bibitem{Sackett2000} Sackett, C. A. {\it et al.}  Experimental entanglement of four particles. \textit{Nature} \textbf{404}, 256--259 (2000).

\bibitem{HVVHsourceGisin}Clausen C. {\it et al.} A source of polarization-entangled photon pairs interfacing quantum memories with telecom photons. \textit{New Journal of Physics} \textbf{16}, 093058(2014).


%\bibitem{FaultTolerant} Lund, A.P., Ralph, T. C. \&  Haselgrove, H. L. Faulttolerant linear optical quantum computing with small-amplitude coherent states. \textit{Phys. Rev. Lett.} \textbf{100}, 030503 (2007).

\bibitem{teleportZeilinger} Bouwmeester, D., Pan, J.-W.,  Mattle, K.,  Eibl, M., Weinfurter, H. \&  Zeilinger A. Experimental quantum teleportation. \textit{Nature} \textbf{390}, 575-579 (1997).

\bibitem{HOM} Hong, C. K., Ou, Z. Y. \&  Mandel, L. Measurement of subpicosecond time intervals between two photons by interference.  \textit{Phys. Rev. Lett.} \textbf{59}, 2044--2046 (1987).

%\bibitem{ReverseHong} Ulanov, A. E., Fedorov, I. A., Sychev, D., Grangier, P. \&  Lvovsky, A. I. Loss-tolerant state engineering for quantum-enhanced metrology via the reverse Hong-Ou-Mandel effect. \textit{Nat. Commun.} \textbf{7}, 11925 (2016).

%\bibitem{Cultivation} Sychev, D. V., Ulanov, A. E., Pushkina, A. A., Richards, M. W., Fedorov, I. A., and Lvovsky, A. I., \textit{Nature Photonics} \textbf{11}, 379–382 (2017). 

\bibitem{Aposterioriteleportation} Braunstein, S. L.\& Kimble, H.J. A posteriori teleportation. \textit{Nature} \textbf{394}, 840--841 (1998)

\bibitem{Pan2003} Pan, J.-W., Gasparoni, S., Aspelmeyer, M., Jennewein, T., \&  Zeilinger, A. Experimental realization of freely propagating teleported qubits. \textit{Nature} {\bf 421}, 721--725 (2003).


\bibitem{Massar1995}  Massar S. \& Popescu S. Optimal Extraction of Information from Finite Quantum Ensembles. \textit{Phys. Rev. Lett.} \textbf{74}, 1259 (1995).

\bibitem{Pan1998} Pan, J.-W., Bouwmeester, D., Weinfurter, H. \& Zeilinger, A. Experimental entanglement swapping: entangling photons that never interacted. \textit{Phys. Rev. Lett.} \textbf{80}, 3891--3894 (1998).

\bibitem{heraldedHVVH1} Barz, S., Cronenberg, G., Zeilinger, A. \&  Walther P. Heralded generation of entangled photon pairs. \textit{Nat. Photon.} \textbf{4}, 553--556 (2010).

\bibitem{heraldedHVVH2}  Wagenknecht, C., Li, C.-M., Reingruber, A., Bao, X.-H., Goebel, A., Chen, Y.-A., Zhang, Q., Chen, K. \& Pan, J.-W. Experimental demonstration of a heralded entanglement source. \textit{Nat. Photon.} \textbf{4}, 549--552 (2010).

\bibitem{heraldedHVVH3} Chen, L.-K. {\it et al.} Experimental nested purification for a linear
optical quantum repeater. \textit{Nat. Photon.} \textbf{11}, 695--699 (2010).

%\bibitem{HVVHsourceSagnac} Kim, T., Fiorentino, M. \& Wong F. N. Phase-stable source of polarization-entangled photons using a polarization sagnac interferometer. \textit{Phys. Rev. A} \textbf{73}, 012316(2006).

%\bibitem{AdvancesQT} S. Pirandola, J. Eisert, C. Weedbrook, A. Furusawa and S. L. Braunstein. Advances in quantum teleportation. \textit{Nature Photon} {\bf 9}, 641 (2015).

\bibitem{Blinov} Blinov, B. B., Moehring, D. L., Duan, L.-M. \& Monroe, C. Observation of entanglement between a single trapped atom and a single photon. \textit{Nature} \textbf{428}, 153--157 (2004).

\bibitem{Kumar2012}Kumar, R. {\it et al.} Versatile wideband balanced detector for quantum optical homodyne tomography. \textit{Opt. Commun.} \textbf{285}, 5259--5267 (2012).

\bibitem{Masalov2017} Masalov, A. V., Kuzhamuratov, A. \&  Lvovsky, A. I. Noise spectra in balanced optical detectors based on transimpedance amplifiers. \textit{Rev. Sci. Instrum.} \textbf{88}, 113109 (2017).

\bibitem{Sackett} Sackett, C. A., {\it et al.} Experimental entanglement of four particles. \textit{Nature} \textbf{404}, 256--259 (2000).

\bibitem{Huisman} Huisman, S. R., {\it et al.}
Instant single-photon Fock state tomography.
\textit{Opt. Lett.} {\bf 34}, 2739--2741 (2009).


\end{thebibliography}
\end{document}